\documentstyle[aps,multicol,prl,psfig]{revtex}
\begin{document}

\draft

\title{Induced coherence with and without induced emission}
\author{H.M. Wiseman$^{1,2}$, K. M\o lmer$^{2}$}
\address{$^{1}$School of Science, Griffith University, Nathan, Brisbane,
Queensland 4111 Australia. \\
$^{2}$Institute of Physics and Astronomy, University of Aarhus,
DK 8000 \AA rhus C, Denmark.}
\maketitle

\begin{abstract}
We analyze signal coherence in the setup of Wang, Zou and Mandel, 
where two optical downconverters have indistinct idler modes. 
Quantum interference, caused by indistinguishability of paths, 
has a visibility proportional to the transmission amplitude between idlers. 
Classical interference, caused by induced emission, may be complete 
for any finite transmission.

\end{abstract}

\newcommand{\beq}{\begin{equation}}
\newcommand{\eeq}{\end{equation}}
\newcommand{\bqa}{\begin{eqnarray}}
\newcommand{\eqa}{\end{eqnarray}}
\newcommand{\nn}{\nonumber}
\newcommand{\nl}[1]{\nn \\ && {#1}\,}
\newcommand{\erf}[1]{Eq.~(\ref{#1})}
\newcommand{\dg}{^\dagger}
\newcommand{\sq}[1]{\left[ {#1} \right]}
\newcommand{\cu}[1]{\left\{ {#1} \right\}}
\newcommand{\ro}[1]{\left( {#1} \right)}
\newcommand{\an}[1]{\left\langle{#1}\right\rangle}
\newcommand{\vl}[1]{\left|{#1}\right|}

\begin{multicols}{2}
The phenomenon of induced emission (that is, emission stimulated in a
system by an input from another system) is well-known in laser
technology \cite{injlock}.  It causes the phase of the amplified field to
adopt the same phase as the incident locking field.
It can also be used in parametric down conversion to lock
the phase of the idler, and hence that of the signal (since the phase
sum of the signal and idler is locked to the pump phase) 
\cite{idlerlock}.
If the field used to lock the
idler of one downconverter (DC2 in Fig.~1) is itself the idler output
of another downconverter (DC1 in Fig.~1), the two signal fields 
will be locked in phase also.
Thus they will have (in principle) perfect first order coherence and so
will interfere at the final beam splitter in Fig.~1. If there
is no connection between the two downconverters, and hence no induced
emission, the two signals wil be
incoherent, and there will be no interference. The 
 classical explanation for this is that 
 in parametric downconversion the phase of the signal 
and idler vary  randomly from shot to shot, with only
their sum being fixed by the pump phase.

The above arguments are completely classical.  Wang, Zou and Mandel
\cite{WanZouMan91}
(WZM) used a completely different (quantum mechanical) explanation,
based on indistinguishability of paths, to explain the interference
they observed in their realization of the experiment shown in Fig.~1.
They did this for the very good reason that there was no induced
emission in their experiment, as the downconversion rates were so low
that the probability of both crystals producing a downconverted pair
was negligible.  Nonetheless, their analysis showed that for perfect 
matching of idler modes, the signal fields from DC1
and DC2 show perfect interference, while the interference is lost
if the idler fields are distinquishable.

The quantum analysis used by WZM is the only
correct explanation of their experiment.  However, the existence of a
classical theory which also reproduces these coherence features 
poses the following question: when is interference due to 
induced emission and when
is it due to indistinguishability of quantum transition paths?
Put another way, what, in the results of WZM,
is the signature
of quantum induced coherence, as distinct from classical induced
emission?
In this letter we show that the signature is the linear
dependence of the coherence on the transmission amplitude $t$ from the
output of idler 1 to the input of idler 2.

Before presenting our analysis, we note that the question of classical 
versus quantum explanations 
for first-order interference in parametric down conversion 
has arizen before \cite{discussions} with reference to an experiment 
of Herzog {\em et al.} \cite{herzog}.  In this experiment
both signal and idler
fields were reflected and passed through a single down-converter
a second time.  Both classical and quantum arguments predict
first-order interference features in the resulting
fields, but also here with different magntudes of 
visibility \cite{discussions}.
An elegant experiment with a single, but spatially extended, 
down-converter was recently performed, where the same discussion
appears as to whether the signal and idler fields 
stimulate down conversion of future pump pulses further along the crystal, 
or whether different pulses interfere because of the indistinguishability
between photons created at different times and places inside the
crystal \cite{kim}.

We turn now to our, fully quantum, analysis of the WZM 
experiment. In an appropriate limit the system can be described by four
modes, $s_1, i_1$ (the signal and idler for DC1), and
$s_2, i_2$ (the signal and idler for DC2).
Consider an arbitrary operator
in the Hilbert space of these four modes.
The equation giving its transformation from its value $O$
before the interaction to its value $O'$ after
the action of the downconverters and the idler transmission 
between DC1 and DC2 is
\beq \label{ut}
O' = U_1\dg U_t\dg U_2\dg O U_2 U_t U_1
\eeq
Here $U_\mu$ for $\mu=1,2$ describe the downconversion in the undepleted
pump
approximation. The crystals and pumps are assumed to
be identical so that
\beq
U_\mu = \exp[-i\chi(a_{s_\mu}a_{i_\mu} + a_{s_\mu}\dg a_{i_\mu}\dg)]
\eeq
where the $a$'s represent annihilation operators. In between the
downconversions the idler from DC1 is put through a beam
splitter, and becomes the idler for DC2. This is described by
\beq
U_t = \exp[(\arcsin t)(a_{i_1}\dg a_{i_2} - a_{i_2}\dg a_{i_1})],
\eeq
where the beam splitter
transmittance $t$ can vary between zero (where idler 2 is
independent from idler 1) and unity (where idler 1 output is equal to
idler 2 input).

Using \erf{ut} we easily obtain the following
\bqa
a_{s_1}' &=& a_{s_1}\cosh \chi - ia_{i_1}\dg \sinh\chi \\
a_{s_2}' &=& a_{s_2}\cosh \chi - ir a_{i_2}\dg\sinh\chi
\nl{-} it
(a_{i_1}\dg \cosh \chi + ia_{s_1}\sinh \chi)\sinh\chi,
\eqa
where $r=\sqrt{1-t^2}$.
Since all of the initial fields are in the vacuum state it is
easy to obtain the expectation values
\bqa
\an{ {a_{s_1}'}\dg a_{s_1}' } &=& \sinh^2\chi \\
\an{ {a_{s_2}'}\dg a_{s_2}' } &=&
\sinh^2\chi \; (r^2+t^2\cosh^2\chi) \\
\an{ {a_{s_1}'}\dg a_{s_2}' } &=& \sinh^2\chi\; t\cosh \chi
\eqa
Note that the two signal modes have equal intensity only in the
limit $\chi \ll 1$.

The maximum obtainable visibility between two fields in an
experiment is given by the modulus of the
first order coherence function
between those fields,
\beq
g^{(1)}(1,2) = \vl{\an{a_1\dg a_2}}
/{\sqrt{\an{a_1\dg a_1}\an{a_2\dg a_2}}}
\eeq
In this case we find between the two final signal fields
\beq \label{g1}
g^{(1)}(1,2) = \frac{ t\cosh\chi}{\sqrt{1+t^2\sinh^2\chi}}.
\eeq
Noting that idler 1, before it enters the beam
splitter, has the same statistics as signal 1, we can rewrite
(\ref{g1}) in terms of the mean photon number $\bar{n}_1 = \sinh^{2}\chi$
entering the beam splitter as
\beq \label{thisform}
g^{(1)}(1,2) = t \sqrt{\frac{1+\bar{n}_1}{1+t^2 \bar{n}_1}}.
\eeq

In this form it is easy to consider the relevant limits.  The
single-photon regime, which is the regime of the experiment and theory
in Ref.~\cite{WanZouMan91}, occurs for $\bar{n}_1 \ll 1$.  That is, the
probability of a
downconversion at DC1 is small, and hence the probability to have
downconversions at both crystals is negligible.  Then the analysis of
WZM applies and we expect the maximum visibility to be equal to
$t$.  This is exactly what \erf{thisform} predicts.

The opposite regime is that where $\bar{n}_1 \gg 1$.  Here there are many
photons on average in all of the downconverted beams.  Thus, we could
expect the classical argument to apply (although our analysis
remains of course completely quantum mechanical).  That is, the phase
of idler 1 should lock the phase of idler 2 for any nonzero
transmittance $t$.  Again, this is reproduced by \erf{thisform}, which in
the limit $\bar{n_1} \to \infty$ is equal to unity for $t>0$ and zero for $t=0$.

For finite values of the first idler photon number output $\bar{n_1}$, the
maximum
visibility is a concave-down function of $t$, as shown in Fig.~2.  It
is evident that even photon numbers of order unity produce marked
deviations from linearity.  This would be interesting to observe
experimentally.

To conclude, we have shown that, regardless of the number of photons
involved, the first-order coherence of two signal beams is unity when
one idler perfectly seeds the second, and zero when the two are
independent.  The difference between the quantum (single-photon) and
classical (many-photon) regimes is for intermediate values of $t$, the
transmittance of the beam splitter which transmits the first idler
output into the second crystal.  A linear dependence of visibility on
$t$, as seen convincingly in Ref.~\cite{WanZouMan91}, is the true
signature of induced coherence without induced emission.

\end{multicols}
\newpage \phantom{.}

\begin{figure}[htbp]
\centerline{\psfig{file=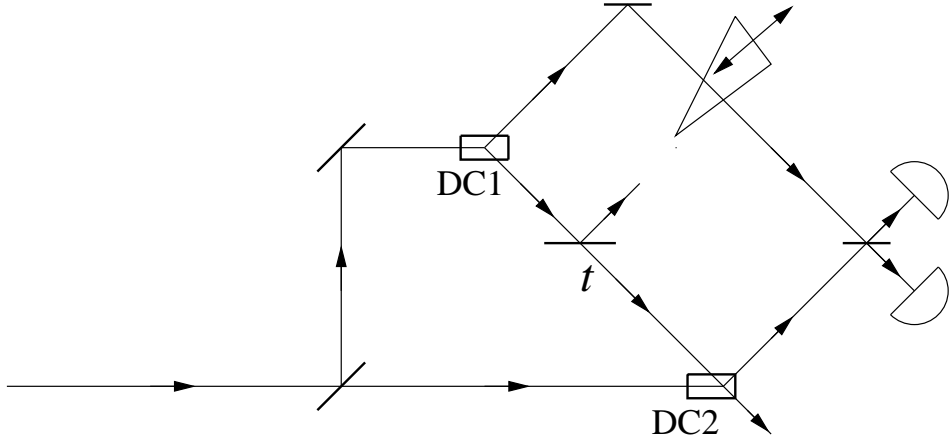,angle=-90}}
\vspace{.5cm}
\caption 
{Experimental setup applied and analyzed by WZM. Two downconverters,
DC1 and DC2,
pumped by a light from a common source are aligned so that the
idler photon from DC1 is injected into DC2
after transmission through a beamsplitter with transmission amplitude
$t$. The signal photons from the downconverters are combined by
another beamsplitter, and an interference signal is recorded as function
of the variation in path length of one of the signal fields.}
\end{figure}

\begin{figure}[htbp]
\centerline{\psfig{file=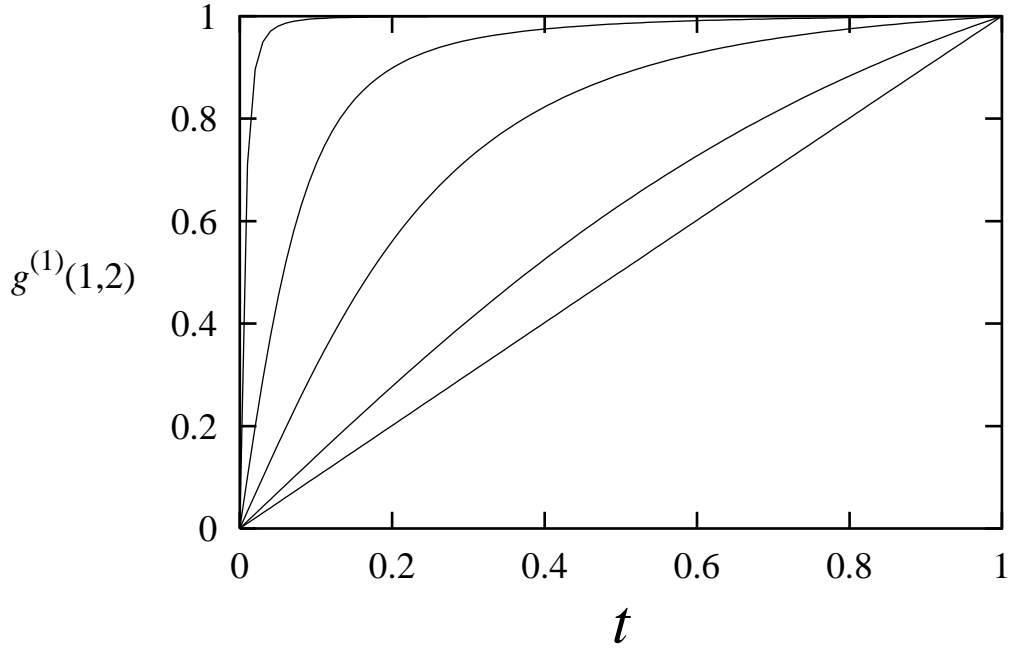,angle=-90}}
\vspace{.5cm}
\caption
{First order mutual coherence function $g^{(1)}(1,2)$  of the
two signal fields, observable as the maximum fringe visibility in the 
interference signal recorded in the set-up of Fig.1. The theoretical
expression (\ref{thisform}) is shown as function of the transmission amplitude
$t$ for different values of $\bar{n}_{1}$, the mean photon number 
entering the beam splitter in Fig.~1. Specifically, from the lowest to 
the highest curve,  
$\bar{n}_{1}=10^{-2}, 1, 10, 100, 10^{4}$.
}
\end{figure}

\begin{references}

\bibitem{injlock}
A.E.Siegman, {\em Lasers} (University
Science, Mill Valley, Calif., 1986); 
H.A.Haus and Y.Yamamoto, Phys. Rev. A {\bf 29}, 1261 (1984); 
T.C.Ralph, C.C.Harb, and H-A Bachor, Phys. Rev. A {\bf 54}, 4359 (1996).

\bibitem{idlerlock}
See for example: M. J. T. Milton, T. D. Gardiner, G. Chourdakis, and P. 
T. Woods, Opt. Lett. {\bf 19}, 281 (1994); R. Urschel {\em et al.},
J. Opt. Soc. Am B{\bf 12}, 726 (1995). 

\bibitem{WanZouMan91}
L.J. Wang, X.Y.~Zou, and L. Mandel,
Phys. Rev. Lett. {\bf 67}, 318 (1991);
X.Y.~Zou, L.J. Wang, and L. Mandel, Phys. Rev. A {\bf 44}, 4614 (1991).

\bibitem{discussions}
A.V.~Belinsky and D.N.~Klyshko, Phys. Lett. A {\bf 166}, 303 (1992);
I.R.~Senitzky, Phys. Rev. Lett. {\bf 73}, 3040 (1994); T.J.~Herzog,
J.G.~Rarity, H.~Weinfurter, and A.~Zeilinger, {\it ibid}, 3041.

\bibitem{herzog} T.J.~Herzog, J.G.~Rarity, H.~Weinfurter, and A.~Zeilinger,
Phys. Rev. Lett. {\bf 72}, 629 (1994).

\bibitem{kim} Y.-H.~Kim, M.V.~Chekhova, S.P.~Kulik, Y.~Shih, and
M.H.~Rubin, quant-ph/9911014.

\end{references}
\end{document}